\documentclass[referee]{aa} 
\usepackage{graphicx}
\newcommand{\harps}{{\textsc{harps}}}

\newcommand{\cible}{{HD\,203608}}

\newcommand{\zeroun}{\delta\nu_{01}}
\newcommand{\ind}[1]{_{\mathrm{#1}}}
\newcommand{\courb}{\mathcal{C}}
\newcommand{\diff}{\mathrm{d}}
\newcommand{\Eqt}{Eq}
\def\m2s2{\,m$^{2}$\,s$^{-2}$} 
\def\kms{\,km\,s$^{-1}$}       
\def\logg{\log g}

\def\niveau{10} 
\newcommand{\nident}{15}  

\begin{document}
\title{HD\,203608, a quiet asteroseismic target in the old galactic disk~\thanks{Based on observations
obtained with the \harps\ \'echelle spectrometer mounted on the 3.6-m telescope at ESO-La Silla Observatory (Chile), programme 077.D-0720}
\thanks{Data corresponding to Fig. \ref{intensity}, Fig. \ref{timeseries} and Table \ref{journal}  are available in electronic form at the CDS via anonymous ftp to cdsarc.u-strasbg.fr (130.79.128.5) or via http://cdsweb.u-strasbg.fr/cgi-bin/qcat?J/A+A/}}
\titlerunning{Asteroseismic study of HD 203608}
\author{
B. Mosser \inst{1}\and
S. Deheuvels \inst{1}\and
E. Michel \inst{1}\and
F. Th\'evenin \inst{2}\and
M.A. Dupret\inst{1}\and
R. Samadi \inst{1}\and
C. Barban\inst{1}\and
M.J. Goupil\inst{1}}

\offprints{B. Mosser}

\institute{LESIA, CNRS, Universit\'e Pierre et Marie Curie, Universit\'e Denis Diderot, Observatoire de Paris, 92195 Meudon cedex, France\\
\email{benoit.mosser@obspm.fr}
\and
Laboratoire Cassiop\'ee, Universit\'e de Nice Sophia Antipolis, Observatoire de la C{\^o}te d'Azur, CNRS, BP 4229, 06304 Nice Cedex 4, France
}
\date{Submitted: April 2007}

\abstract{A short observing run with the spectrometer \harps\ at the ESO 3.6-m telescope was conducted in order to continue exploring the asteroseismic properties of F type stars. In fact, Doppler observations of F type on the main sequence are demanding and remain currently limited to a single case (HD 49933). Comparison with photometric results obtained with the CoRoT mission on similar stars will be possible with an enhanced set of observations.}%
{We selected the 4th magnitude F8V star \cible, in order to investigate the oscillating properties of a low-metallicity star of the old galactic disk.}%
{A 5-night asteroseismic observation program has been conducted in August 2006 with \harps. Spectra were reduced with the on-line data reduction software provided by the instrument. A new statistical approach has been developed for extracting the significant peaks in the Fourier domain.}%
{The oscillation spectrum shows a significant excess power in the frequency range [1.5, 3.0 mHz]. It exhibits a large spacing about 120.4\,$\mu$Hz at 2.5 mHz. Variations of the large spacing with frequency are clearly identified, which require an adapted asymptotic development.
The modes identification is based on the unambiguous signature of \nident\  modes with $\ell = 0$ and 1.}%
{This observation shows the potential diagnostic of asteroseismic constraints. Including them in the stellar modeling  enhances significantly the precision on the physical parameters of \cible, resulting in a much more precise position in the HR diagram. The age of the star is now determined in the range $7.25\pm0.07$\,Gyr.}
\maketitle
\section{Introduction}

New stable spectrometers dedicated to very precise radial velocity measurements have permitted rapid progress in observing solar-like oscillations in solar like stars (see for example \cite{2006ESASP.624E..25B} for a recent review). The questions raised with the observation of the CoRoT target HD49933 (\cite{2005A&A...431L..13M}), an active F5V star, led us to continue the observations of F stars. We therefore planned to measure, identify and characterize the solar-like oscillations of another low-metallicity F type dwarf star, with a 5-night run.

The selected target \cible\ (HIP 105858, HR 8181, $\gamma$ Pav) is a F8V star
which belongs to the group of the Vega-like stars with an infrared excess attributed to the presence of circumstellar dust warmed by the central star. 
Its age is estimated to range between 6.5 to 14.5 Gyr (\cite{2006ApJ...636.1098B}), according to previous work including \cite{1999A&A...348..897L}, based on different estimators: comparison to theoretical isochrones, rotational velocity, strength of chromospheric calcium
emission lines, stellar metallicity, and space velocity. Its metallicity [Fe/H]=$-$0.65 dex designates it as an old dwarf of the optical thick disk  according to \cite{1993A&A...275..101E}, but \cite{2006MNRAS.367.1181B} describes it as a member of the thin disk.

\cite{2003A&A...400..241P} has already identified \cible\ as a possible asteroseismic target. Its right ascension and declination are favorable for a
single-site campaign at La Silla during winter. Moreover this star has a low projected rotational velocity ($v\sin i=2.4 \pm 0.5$\kms, \cite{2003A&A...398..647R}) which is excellent for Doppler asteroseismology, especially for a F star. On the other hand, with broader lines than G stars, F stars are demanding targets for ground-based seismic Doppler observations.
As a result, \cible\ is an exciting star to compare with a similar star, the active F5V star HD 49933, a main target of the satellite CoRoT already studied by spectrovelocimetry (\cite{2005A&A...431L..13M}).

Section~\ref{etoile} reports the current status on the physical parameters of \cible.
Observations are presented in Section~\ref{observations}, with the analysis of the time series and of the activity signal.
The seismic analysis, based on the unambiguous detection of $\ell$ = 0 and 1 modes is exposed in Section~\ref{signature}. Asymptotic parameters are extracted from the Fourier spectrum, then individual eigenfrequencies and amplitudes.
The modeling of \cible\ is presented in Sect.~\ref{modeling}. Section~\ref{conclusion} is devoted to conclusions.


\section{Stellar parameters\label{etoile}}

\subsection{Temperature and luminosity}

The atmospheric parameters of \cible\ have been discussed in several studies (see \cite{1997A&AS..124..299C}). It results a range of 5929\,K $<T\ind{eff}<$ 6139\,K
for the effective temperature. Using photometric data, \cite{1998yCat.3193....0T} and more recently \cite{2005A&A...440..321J} and \cite{2005A&A...441.1149D} adopted a similar value: $T\ind{eff}=6070$\,K with an uncertainty of $\pm$100\,K. But this star is a Vega-like one which exhibits infrared excess (\cite{1998ApJ...497..330M}). The flux absorption and reemission in the infrared is difficult to estimate, and its effective temperature may be underestimated.

The absolute bolometric magnitude was determined using the apparent V magnitude (Table \ref{prop-radius}) combined with its Hipparcos parallax ($\pi=107.98 \pm 0.19$ mas, \cite{2007hnrr.book.....V}). The bolometric correction used is from \cite{1996ApJ...469..355F}. The resulting luminosity is $L/L_\odot=1.39 \pm 0.13$ (Table \ref{prop-phys}).

\subsection{Abundances}

The atmospheric stellar abundance published in the recent literature converges
to the value $-$0.67 dex for the iron element with a detailed analysis under LTE assumption.
Using NLTE computations, \cite{1999ApJ...521..753T} (hereafter IT99) have proposed to correct both this iron LTE abundance and the spectroscopic surface gravity.
They have shown that the more metal-poor the star, the more important NLTE effects:
the ionization balance of iron adopted to determine the spectroscopic $\logg$
is in error. IT99 having adopted $T\ind{eff}=6072$\, K. Like \cite{2005A&A...440..321J},
we can use the NLTE values proposed by IT99 for [Fe/H]:
$-$0.57\,dex. The $\logg$ value is deduced from the Hipparcos distance. This parameter
has a large uncertainty ($\simeq$0.15\,dex) due mainly to the bolometric correction, and therefore cannot be used  to constrain efficiently the stellar fundamental parameters.  \cite{2003A&A...407..691K} found also NLTE effects on the lines of the iron element in metal-poor stellar atmospheres
but in a lesser quantity since they adopted for the NLTE computations
a higher efficiency of the inelastic collisions with hydrogen. In the absence of a realistic
theory of inelastic collisions with H, we shall adopt a NLTE abundance of  $-0.60$\,dex for Fe. In most of metal-poor stars the alpha elements present an enhancement compared to
iron element abundance, up to 0.4\,dex. \cite{2000ApJ...541..207I} have
computed NLTE abundances for Ca and Mg elements in \cible. For that star,
Ca appears to follow Fe, contrary to Mg which is slightly enhanced compared to Fe.
In the LTE abundance analyses, \cite{1998yCat.3193....0T} did not found any strong enhancement of the alpha elements Ca, Mg, Si, and only a slight enhancement for the oxygen element:
[O/Fe]$\simeq$0.1. Then, we shall adopt a moderate [(alpha elements + O)/Fe] value, about 0.1. It results that we adopt for the modeling of \cible\ a surface metallicity of [Z/X]$\ind{f}=-0.50\pm 0.10$ with respect to the solar metallicity (\cite{1993PhST...47..133G}). Because this star belongs to the old disk, we shall adopt for the calibration an initial abundance Y$\ind{i}$=0.25.

\begin{table}
\caption{Color magnitudes of \cible, and inferred radii. }\label{prop-radius}
\begin{tabular}{ccr}
\hline
Band&Magnitude&$R/R_\odot$\\
\hline
V & 4.22 & 0.96$\pm$0.08 \\
R & 3.75 & 1.06$\pm$0.06\\
I & 3.45 & 1.08$\pm$0.05\\
J & 3.27 & 1.04$\pm$0.04\\
K & 2.90 & 1.08$\pm$0.02\\
L & 2.85 & 1.09$\pm$0.02\\
\hline
\end{tabular}\end{table}

\begin{table}
\caption{Physical parameters of \cible}\label{prop-phys}
\begin{tabular}{lr}
\hline
$T\ind{eff} (K)$   & 6070$\pm$100\\
$[$Z/X$]\ind{f}$     & $-0.50\pm0.10$\\
$L/L_\odot$   & 1.39$\pm$0.13\\
$\Pi$   (mas)  & 107.98$\pm$0.19\\
$\log g$ (cm\,s$^{-2}$)    & 4.30$\pm$0.15  \\
$v \sin i$ (\kms) & 2.4$\pm$0.5\\
%
\hline
\end{tabular}\end{table}

\begin{figure}
\centering
\includegraphics[width=8.5cm]{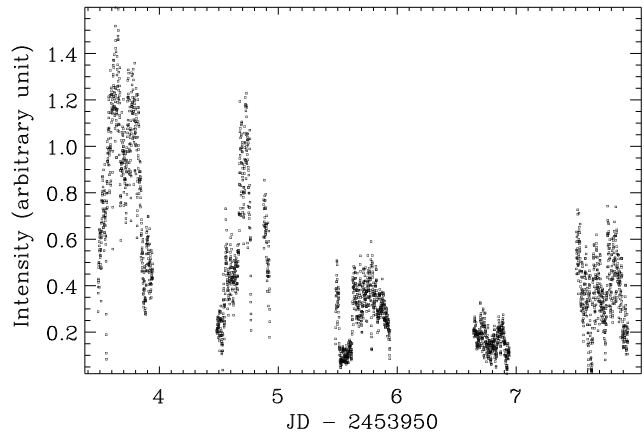}
\caption{Time series of the signal intensity. Night 4 suffered from an unusually large seeing (up to 3.5"). The efficiency of the collected signal is highly sensitive to the sky opacity and to the seeing.}
\label{intensity}
\end{figure}

\begin{figure}
\centering
\includegraphics[width=8.5cm]{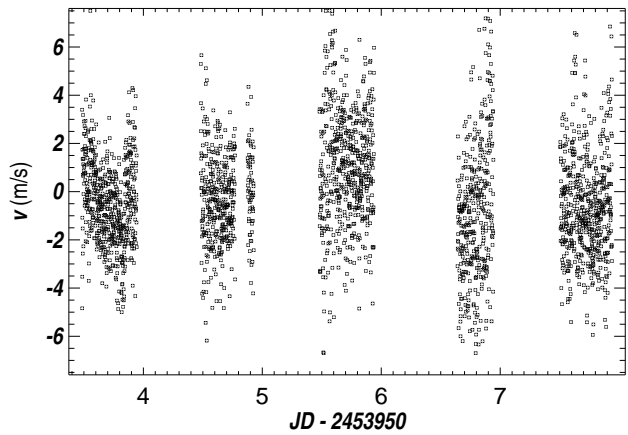}
\caption{Time series of the radial velocity of \cible\ measured with the pipeline reduction of \harps\ (unfiltered data). The signal is free of any important low-frequency component.
\label{timeseries}}
\end{figure}

\begin{figure}
\centering
\includegraphics[width=8.5cm]{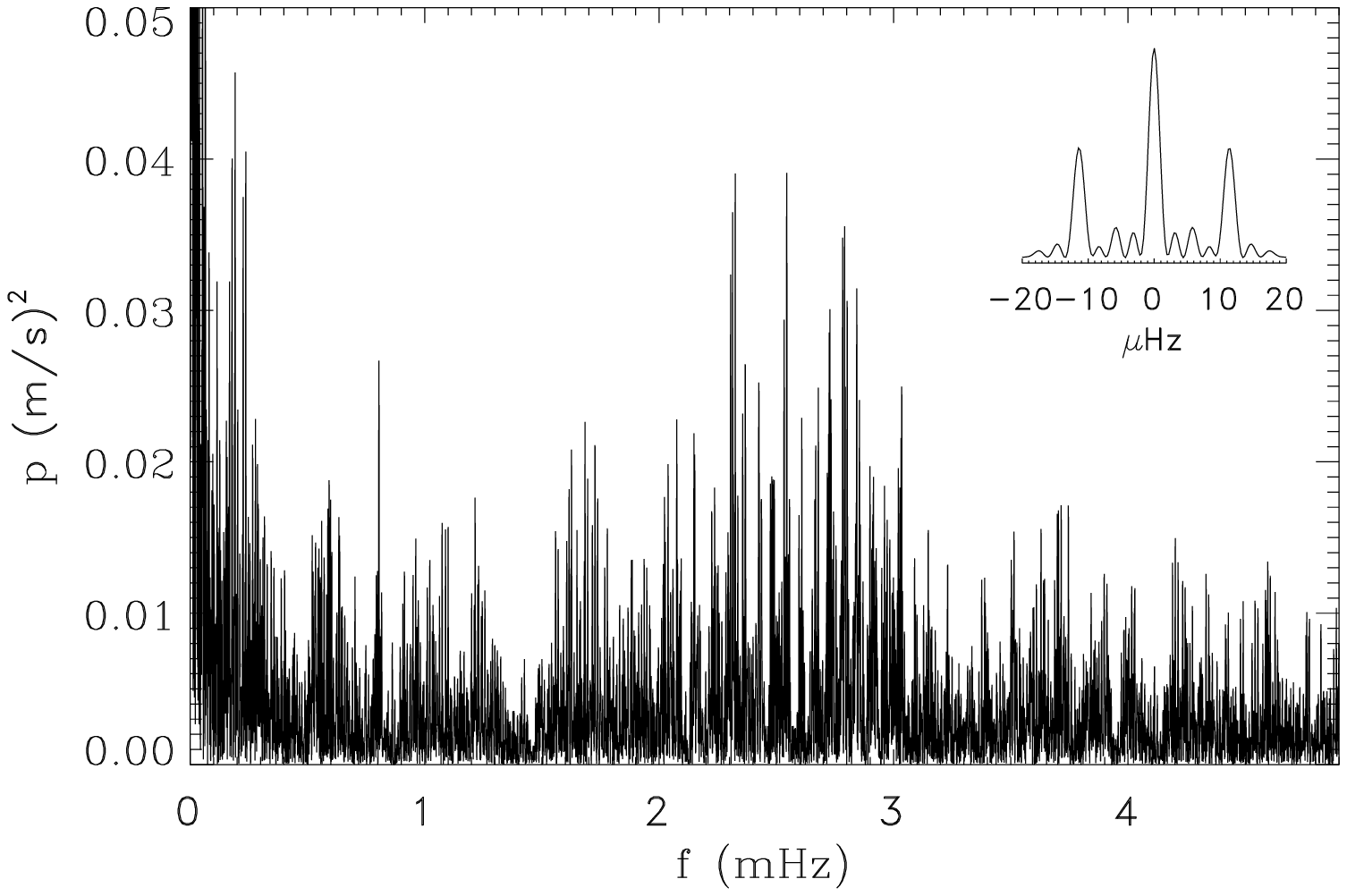}
\caption{Lomb-Scargle periodogram, and inset of the window function. The time series used for this spectrum excludes the noisiest value; night 4 is completely filtered out with such a treatment. 
\label{spectrum}}
\end{figure}

\subsection{Mass and radius}

The accurate Hipparcos parallax determination 
may help to constrain the adopted $T\ind{eff}$ value with the estimate of the radius derived from surface brightness relations, as already done for several asteroseismic studies (\cite{2004A&A...426..297K}, \cite{2005A&A...436..253T}). Values of the radius derived from the magnitudes in different infrared bands are given in Table~\ref{prop-radius}. \cite{2007ApJ...659..616C} have demonstrated the advantage to constrain the stellar modeling  with asteroseismic data with an independent interferometric measure of the radius.
However, the  infrared excess and the presence of the disk around the star make this estimate inaccurate, as shown for the Vega-like star $\tau$ Ceti. Its diameter measured with interferometry has been first used to constrain its age and mass (\cite{2004A&A...426..601D});  but recently, \cite{2007arXiv0710.1731D} have measured  the K band excess due to hot dust exozodiacal disk, and then found a smaller diameter. Therefore, we prefer not to use in this study the \cible\ diameter as a fixed constraint.

The mass of \cible\ has been estimated by \cite{2005A&A...440..321J} using evolutionary tracks inversion and assuming an age of 14\,Gyr :  $M/ M_{\odot} = 0.88$. No uncertainty on the mass has been proposed by the authors. As for the radius, we cannot consider the stellar mass as a constraint. On the other hand, one of the objective of these observations is to show the capability of asteroseismology to derive accurate values of these parameters.

\section{Observations\label{observations}}

About 47.6 hours observations were obtained between August 5 and 10, 2006, representing 2504 individual measurements (Table \ref{journal}). The exposure time was 33 s, giving one measurement each 68.5 s, and a Nyquist frequency about 7.3 mHz, much above the predicted oscillation cutoff frequency.
For 4 nights out of 5, the observation duration at this winter period was greater than 10.3 hours per night. The fourth night was affected by strong perturbations, with an extremely bad seeing (up to 3.5") and many cirrus. Both effects yielded a low intensity signal (Fig.~\ref{intensity}): most of the photons do not enter the 1" fibre when the seeing is too much degraded.

The mean SNR in the \'echelle spectrum at 550 nm was typically better than 250 at airmass less than 2. The noise level finally derived from the high frequency part of the Fourier spectrum is about 5.1\,cm\,s$^{-1}$, corresponding to a high frequency noise in the time series about 1.6\,m\,s$^{-1}$ and to a noise equivalent velocity of about 1.0\,m\,s$^{-1}/\smash{\sqrt{\hbox{mHz}}}$.

\begin{table}
\caption{Journal of radial velocity measurements. The 2nd night was split in two parts. $\Delta T$ represents the length of observation each night.
The dispersion $\sigma\ind{RV}$ is derived from the high frequency noise recorded in the power spectrum of each night. \label{journal}}
\begin{center}
\begin{tabular}{cccc} \hline Date & Number of & $\Delta T$ & $\sigma\ind{RV}$  \\
Aug. 06 & spectra & (hr) &  (m\,s$^{-1}$) \\
\hline

05 & 583  & 11.0 &  1.5 \\
06 & 358 +65 &  6.9 + 1.2 &  1.7 + 1.7 \\
07 & 570  & 10.9 &  2.6\\
08 & 377  &  7.2 &  3.0 \\
09 & 551  & 10.4 &  2.2\\
\hline
\end{tabular}
\end{center}
\end{table}

The time series show that  \cible\ does not in fact present any signature of activity (Fig.~\ref{timeseries}), contrary to HD 49933 that showed a strong low-frequency modulation (\cite{2005A&A...431L..13M}). This difference can be due to the higher stellar age. It can be also related to a low value of the stellar inclination, accounting for the low $v\sin i$ value. The quasi pole-on observations should then hamper any significant signature of the rotational modulation.

\begin{table}
\caption{Different treatments of the time series, with different filter levels (arbitrary unit, similar to the y-axis in Fig.~\ref{intensity}). The dispersion $\sigma_\nu$ is directly measured in the high frequency part of the power spectrum; $\sigma_t = \sigma_\nu \sqrt{n\ind{c}/2}$ is the high frequency noise in the time series, with $n\ind{c}$ the number of points in the time series.
}\label{traitements}
\begin{tabular}{rrrrr}
\hline
cut  & $\eta$ & $n\ind{c}$ &  $\sigma_t$& $\sigma_\nu$  \\
level     & \%&            &  (m\,s$^{-1}$)       & (cm\,s$^{-1}$)          \\
\hline
  0 & 45 & 2054  &  2.0  &  5.7  \\
0.1 & 42 & 2373  &  1.8  &  5.3  \\
0.2 & 35 & 1960  &  1.6  &  5.1  \\
0.3 & 28 & 1580  &  1.4  &  5.1  \\
0.4 & 21 & 1165  &  1.4  &  5.7  \\
0.5 & 14 &  811  &  1.3  &  6.2  \\
\hline
\end{tabular}\end{table}

\section{The seismic signature\label{signature}}

\subsection{Power spectra}

Computing the power spectrum of the complete time series requires a careful treatment of the noisiest data. Different ways have been proposed, as the weights introduced in the Lomb Scargle analysis by \cite{2004ApJ...600L..75B}. In fact, this method makes the noise uniform in the time series and minimizes the noise level in the power spectrum. As a consequence, it modifies the signal and yields a degraded window function, which increases the interferences in the power spectrum between the signal, the aliases and the noise. 

On the other hand, eliminating noisy data requires a criterion for defining a threshold, and translates  immediately in a degradation of the duty cycle. Optimizing simultaneously the signal-to-noise ratio (SNR) and the duty cycle is therefore necessary. Choosing the optimum threshold for the lowest noise level is possible, but it does not insure the highest efficiency for detecting eigenmodes, since highly unpredictable interferences occur between the signal, its aliases and the noise.

\begin{figure}
\centering
\includegraphics[width=8.5cm]{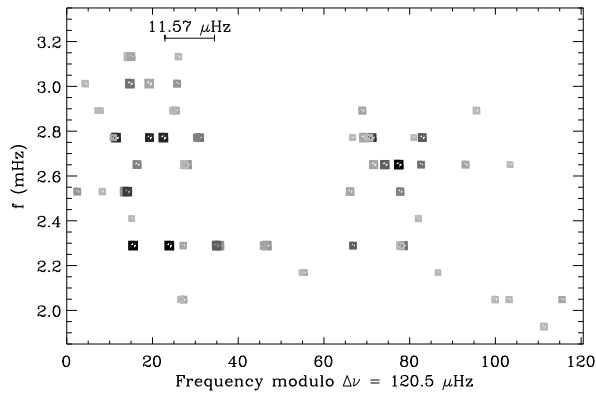}
\caption{\'Echelle diagram, with all peaks with a false alarm probability less than \niveau\%, collected for 10 different treatments (with increasing cutoff in SNR, hence decreasing value of the duty cycle). The size of the symbol is representative of the mode amplitude. A dark symbol indicates a peak present in many treatments, whereas a light grey symbol corresponds to a peak only present in one treatment. The strong peaks at 2.3\,mHz with an abscissa around 40\,$\mu$Hz emerge only in the noisiest data set, and therefore are not selected.
\label{echelle}}
\end{figure}

\begin{figure}
\centering
\includegraphics[width=8.5cm]{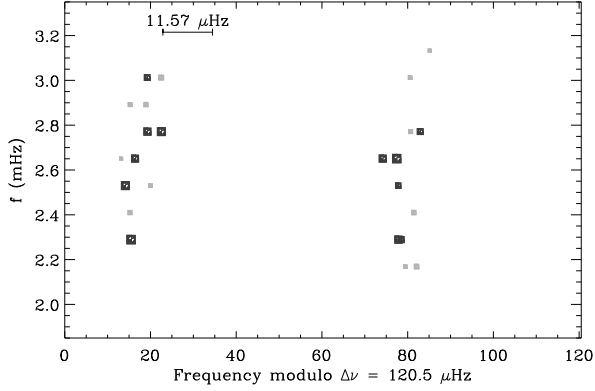}
\caption{\'Echelle diagram, selecting the major peaks detected in the two major ridges identified in Fig.~\ref{echelle} (black symbols), completed with peaks with a larger false alarm probability (up to 25\%, grey symbols) but following the asymptotic pattern, and excluding the peaks present in the neighbor alias of the most prominent ridges.
\label{echelle2}}
\end{figure}

We have endorsed this impossibility of an a-priori criterion for the best solution, and proposed to circumvent it: instead of performing only one single power spectrum, we have calculated many, each one corresponding to a different threshold level in the time series (Table~\ref{traitements}). Thresholds were based on the signal intensity (Fig.~\ref{intensity}).
In each case, the LS periodogram was computed and the highest  peaks were selected according to the analysis in \cite{2004A&A...428.1039A} that gives a test for detecting peaks of short-lived p modes embedded in a power spectrum of noise. Simulations have confirmed that the single treatment corresponding to the optimum solution is far from providing all significant peaks. 
The peaks with a false alarm probability lower than \niveau\% are plotted in an \'echelle diagram based on a large splitting of 120.5~$\mu$Hz (Fig.~\ref{echelle}). The procedure makes then possible the identification of the major ridges in the \'echelle diagram. The most confident peaks in these ridges are then identified (Fig.~\ref{echelle2}); those selected have the minimum false alarm probabilities obtained in each periodogram.

\begin{figure}
\centering
\includegraphics[width=8.5cm]{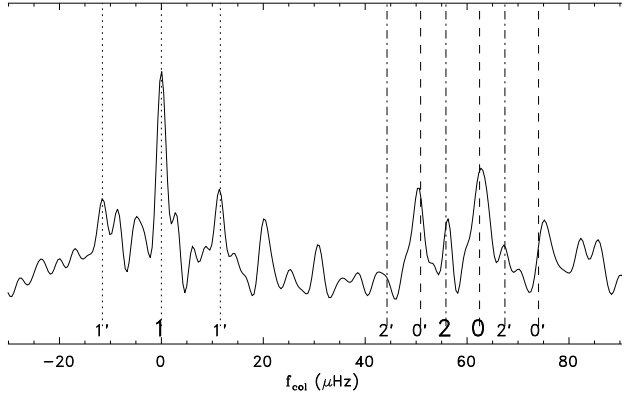}
\caption{Collapsogram of the \'echelle diagram, with all identified large spacings rescaled on one single value $\Delta\nu\ind{eq}$, for modes identified in the range [2.2, 2.9 mHz]. Modes $\ell=0$ or 1 can be identified, as well as their alias (with a prime). The small spacing derived from the frequency difference $\zeroun$ allows us to identify the signature of $\ell=2$ modes.
\label{collapsogram}}
\end{figure}

\subsection{Asymptotic parameters}

The \'echelle diagram exhibits clearly the regular pattern constructed by the asymptotic behavior of low-degree high-frequency pressure modes. Clear ridges appear in the frequency range [2.3 - 2.8 mHz]. The multiple calculations of periodograms, as many as threshold values were fixed, allowed us to put in evidence more peaks than obtained with the calculation of a single spectrum. For 17 peaks (including the aliases) finally detected in the range [2.3 - 2.8 mHz] with a false alarm probability less than 10\%, 12 at most were present in a single power spectrum.

The ridges in the \'echelle-diagram show a curvature in the range [2.3 - 2.8 mHz]. This may be  interpreted as a significant second order term to the asymptotic law, or as the signature of a modulation in the spectrum due to an important density gradient inside the star. In order to account for it, we propose a fit of the frequency pattern varying as:
\begin{equation}
\nu_{n,\ell} = \nu_{n_0} + n_\ell \ \Delta\nu
- \ell (\ell + 1) \ D_0
+ {n_\ell^2 \over 2}\ \courb
\label{developpement}
\end{equation}
with
\begin{equation}n_\ell = n -n_0 + {\ell\over 2}
\label{ordreradial}
\end{equation}
The frequency $D_0$ measures the small spacing, as usually done in most of the previous works reporting single-site asteroseismic observations. $\courb$ represents the variation factor of the large spacing with the radial order, and defines the local curvature in the \'echelle diagram. It corresponds to a global linear increase of the large spacing with the radial order $n$ such as:
\begin{equation}
\Delta\nu (n) = \Delta\nu (n_0) + \courb\, n_\ell
\end{equation}

\begin{figure}
\centering
\includegraphics[width=8.5cm]{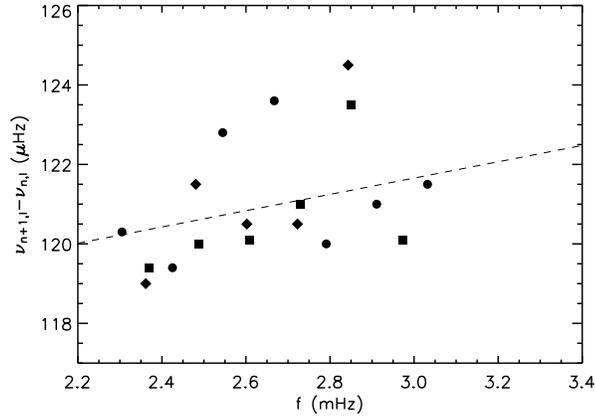}
\caption{Variation of the large spacing with frequency, for the degrees $\ell=0$ (squares), 1 (circles) and 2 (diamonds). The dispersion with respect to the linear fit (dashed line) is compatible with the frequency uncertainty around 1.2\,$\mu$Hz derived from \cite{1992ApJ...387..712L}, that yields twice this uncertainty on each frequency difference.
\label{lspacing}}
\end{figure}

In order to account for the curvature and the irregularities in the \'echelle diagram, we built a rectified collapsogram, as in \cite{2008A&A...478..197M}.
This rectified collapsogram of the \'echelle diagram (Fig.~\ref{collapsogram}), corrected from the variation with frequency of the large spacings between modes of same degree, puts in evidence the signature of $\ell$=2 modes. Their location compared to radial modes is in agreement with the small spacing derived from the frequency difference $\zeroun = \nu_{n,0} - (\nu_{n,1}+\nu_{n-1,1})/2$ derived from $\ell$=0 and 1 modes.
We may then propose an identification of eigenmodes with degrees $\ell=0$, 1 and 2, from which we can derive the local values of the large spacing (Fig.~\ref{lspacing}).

The large spacing is about 120.3$\pm$0.5\,$\mu$Hz at 2.6 mHz, matching the value 119\,$\mu$Hz expected from the scaling based on the square root of the mean density $\smash{\sqrt{{\cal G}M/R^3}}$.
The variation factor $\courb$, around $0.4^{+0.4}_{-0.3}\, \mu$Hz in the range [2.3 - 2.8 mHz], gives a  measure of the variation of the large spacing with the radial order $n$.

\subsection{Small spacings}

\begin{figure}
\centering
\includegraphics[width=8.5cm]{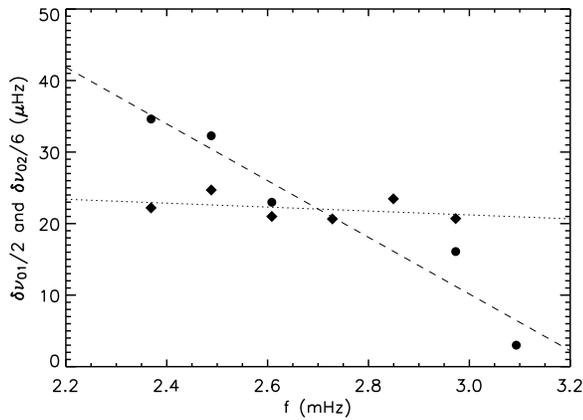}
\caption{Variation of the small spacings $(n+\ell/2)\,\delta\nu_{01}/2$ and $(n+\ell/2)\,\delta\nu_{02}/6$ with frequency,
derived from the $\ell=1$ (circles) and 2 (diamonds) modes, compared to $\ell=0$.
\label{sspacing}}
\end{figure}

Variations of the small spacings with frequency show a more complicated pattern than expected from simple asymptotics (Fig.~\ref{sspacing}). The decrease with frequency of the small frequency differences is so pronounced that, instead of the development introduced with \Eqt~\ref{developpement}, we prefer a development closer to the asymptotic form (\cite {1980ApJS...43..469T}):
\begin{equation}
\nu_{n,\ell} = \nu_{n_0} + n_\ell \ \Delta\nu
- {\ell (\ell + 1)\over n+{\ell/2}}  \ A_0
+ {n_\ell^2 \over 2}\ \courb
\label{tassoul}
\end{equation}
The characteristic frequency $A_0$ introduced in this asymptotic development, around 24$\pm$18\,$\mu$Hz, corresponds at 2.6 mHz to a classical small spacing value $D_0$ about 1.2$\pm$0.9\,$\mu$Hz. The correction introduced with the $1/n$ term appears efficient, but is not sufficient to account for the variation of the small frequency difference. 

The evolved star \cible\ may present a core mainly composed of helium, responsible for a significant central decrease of the sound speed. Such a core has a strong influence on the second order asymptotic term, through the integral term $\diff c / r$, that the classical Tassoul development cannot take into account. The rapid density and sound speed variations may be modeled as a discontinuity. The asymptotic form in that case (\cite{1993A&A...274..595P}, see their \Eqt~4~.1)  introduces many parameters for describing the modulation of the small spacings. With eigenfrequencies identified over a limited frequency range, it has little sense to try to fit all of them, so that we prefer to linearize the development in the form:
\begin{equation}
\nu_{n,\ell} = \nu_{n_0} + n_\ell \ \Delta\nu
- {\ell (\ell + 1)\over n+{\ell/2}}  \ [P_\ell +Q_\ell \  n_\ell]
+ {n_\ell^2 \over 2}\ \courb
\label{provost}
\end{equation}
It yields small spacings such as:
\begin{eqnarray}
\delta\nu_{01}& \simeq & {2\over n}\ [P_1 + Q_1\ n_\ell] - {1\over 8}\ \courb \label{provost01} \\
\delta\nu_{02}& = & {6\over n}\ [P_2 +  Q_2\ n_\ell] \label{provost02}
\end{eqnarray}
with $\delta\nu_{02}= \nu_{n,0}-\nu_{n-1, 2}$.
The parameters introduced by \Eqt~\ref{provost} are summarized in Tab.~\ref{prop-seism}. Compared to the $A_0$ factor introduced by \Eqt~\ref{tassoul}, the relative errors on the parameters $P_1$ and $P_2$ are sensitively reduced. The identification of individual eigenmodes is finally given in Table \ref{identification}. We note that the location of $\ell=2$ modes is possibly influenced by aliases of the $\ell=0$ modes, since the configuration of the time series yield reinforced signatures of the alias related to a 4-day periodicity.

\begin{table}
\caption{Estimation of the asymptotic parameters, relying on the modes detected in the frequency range [2.3, 2.8 mHz], and 3-$\sigma$ error bars.
}\label{prop-seism}
\begin{tabular}{lr}
\hline
Asymptotic parameters (at 2.6 mHz)\\
with a Tassoul-like development\\
\hline
$\Delta\nu$     & 120.3$\pm$0.5 $\mu$Hz\\  
$\courb= \diff \Delta\nu / \diff n$ & $0.4^{+0.4}_{-0.3}\, \mu$Hz \\ 
$A_0$           & 24$\pm$18 $\mu$Hz\\  
$D_0 \simeq A_0 /20$ & 1.2$\pm$0.9 $\mu$Hz\\
\hline
Second order terms\\
\hline
$P_1$  & $26.0 \pm 7.5$ $\mu$Hz\\  
$P_2$  & $22.2 \pm 3.5$ $\mu$Hz\\  
$Q_1$  & $-5.2 \pm 3.2$ $\mu$Hz\\  
$Q_2$  & $-0.3 \pm 1.8$ $\mu$Hz\\  
\hline
Amplitudes & \\
\hline
$\nu\ind{max}$ & 2.6 mHz\\
$v\ind{max}$   & 22$\pm$2\,cm\,s$^{-1}$\\
\hline
\end{tabular}\end{table}

\begin{table}
\caption{Identified peaks, with $\cal P$ their confidence level (here complementary to the false alarm probability), and inferred peaks, as the function of the assumed radial order $n'$. Frequency uncertainty is around 1.2 $\mu$Hz, according to the estimated lifetime, the observation duration and the SNR (\cite{1992ApJ...387..712L}). The detections and identifications outside the range [2.3-2.8 mHz] are not as certain as within it. Confidence levels are given when modes are directly identified in the \'echelle spectrum. For $\ell=2$ modes, they cannot be given, since those modes are just inferred assuming a regular \'echelle pattern.
\label{identification}}
\begin{tabular}{llllll}
\hline
    &\multicolumn{2}{c}{$\ell=0$} &\multicolumn{2}{c}{$\ell=1$}&\multicolumn{1}{c}{$\ell=2$} \\
$n'$ & $\nu\ind{obs}$ & $\cal P$  & $\nu\ind{obs}$ & $\cal P$ & $\nu\ind{obs}$  \\
    &   mHz & \% &   mHz & \% &   mHz  \\
\hline
 17 &       &      & 2.305 &   95 & 2.361\\
 18 & 2.369 &   91 & 2.425 &   87 & 2.480\\
 19 & 2.488 &   79 & 2.545 &   94 & 2.602\\
 20 & 2.608 &   86 & 2.667 &   92 & 2.722\\
 21 & 2.728 &   94 & 2.791 &   93 & 2.843\\
 22 & 2.849 &   88 & 2.911 &   81 & 2.967\\
 23 & 2.973 &   71 & 3.032 &   84 & \\
 24 & 3.093 &   75 & 3.154 &   71 & \\
\hline
\end{tabular}\end{table}

\subsection{Amplitudes}

According to its mass and luminosity and following the power law given by \cite{2007A&A...463..297S}, the maximum amplitudes of \cible\ was supposed to be about 30 cm\,s$^{-1}$. In order to estimate the maximum oscillation amplitude, we have constructed  synthetic time series, based on a theoretical low degree p-modes eigenfrequency pattern. The modes lifetimes are estimated from the eigenfrequency widths, between 1 and 4 $\mu$Hz FWHM (\cite{1999A&A...351..582H}). Due to the short duration of the time series, possible large uncertainties in the lifetimes estimate translate in very small uncertainty in the result.
The maximum amplitudes are assumed to follow a Gaussian distribution in frequency. The synthetic time series are then calculated using the model of a stochastically excited, damped harmonic oscillator (\cite{1990ApJ...364..699A}), and include also a white noise.

The maximum amplitude is adjusted in order to obtain comparable energy per frequency bin in the synthetic and observed spectra (the reference observed spectrum has a cut level at 0.2, what corresponds to the minimum dispersion $\sigma_\nu$, as shown in Table \ref{traitements}). A Monte-Carlo approach finally shows that the best agreement is for a signal with a maximum amplitude about 22$\pm$2\, cm\,s$^{-1}$ rms, with a Gaussian envelope centered at 2.6 mHz and with a 1.2 mHz FWHM. The simulation shows that the noise component, precisely determined due to the large oversampling of the time series, is 1.57$\pm$0.10\,m\,s$^{-1}$, in agreement with the high frequency noise directly determined in the spectrum.

The observed maximum amplitude and the predicted scaling (33$\pm$5 cm\,s$^{-1}$ rms) agree only marginally. The difference may be related to the low metallicity of \cible;  a low metallicity yields a thinner convective envelope, then possibly lower amplitudes. Preliminary 3-D simulations of the outer layers of a star with a metallic abundance 10 times smaller than solar result in a mode excitation rate about 2 times smaller. Actually, low metallicity corresponds to a mean opacity smaller compared to the solar one. Then, in the super-adiabatic region, where convection is inefficient because of the optically thin atmosphere, the radiative flux is larger than in a medium with a solar metallicity. In that case, convection can be less vigorous for evacuating the same amount of energy, leading to a weaker driving. Therefore, it seems coherent that p modes in a star with sub-solar metallicity [Fe/H]$\simeq -$0.6 dex present significantly smaller amplitudes than in a star with solar metallicity. 

This preliminary analysis has to be refined, since previous asteroseismic subgiant targets with much lower or greater metallicity than the Sun did not show such a discrepancy compared to \cite{2007A&A...463..297S} empirical law ($\nu$ Ind, [Fe/H]$\simeq -1.4$, \cite{2007A&A...470.1059C}; $\mu$ Ara, [Fe/H]$\simeq$0.35, \cite{2005A&A...440..615B}). On the contrary, the dwarf target HD 49933 ([Fe/H]$\simeq -0.37$, \cite{2005A&A...431L..13M}) already showed weaker amplitudes than expected.

\section{Modeling\label{modeling}}

Interior models taking into account the new asterosismic constraints were all computed using the evolutionary code CESAM2k (\cite{1997A&AS..124..597M}).
We used the OPAL equation of state and the nuclear data from NACRE (\cite{1999NuPhA.656....3A}). The boundary layers are described using a model of atmosphere derived from the Kurucz model 
adapted to an undermetallic star (\cite {1997ESASP.419..193K}). For the chemical composition, we used the solar mixture from Grevesse \& Noels (1993). The revised abundances from \cite{2005ASPC..336...25A} suggest lower abundances of C, N, O, Ne and Ar. Many studies showed however that the standard solar models using the new abundances of Asplund are in disagreement with the sound speed profile, the radius and the helium abundance of the
convection zone (\cite{2005ApJ...627.1049G}).
These new abundances need to be confirmed or infirmed by independent 3D NLTE line transfer studies of the oxygen element. Nevertheless, we also computed models with the abundances of Asplund, and checked that they do not induce any significant change in the results. The convection follows the description of \cite{1991ApJ...370..295C} with a mixing length parameter $\lambda$ close to 1. A model of the Sun using this description of the convection led to $\lambda \simeq 0.94$. This value was then adopted for this study. The stellar models were computed with microscopic diffusion using the formalism developed by \cite{1969fecg.book.....B}.

\subsection{Description of the fit}

The models are constrained using the following observational quantities : $T\ind{eff}$,
$L/L_{\odot}$, [Z/X$]\ind{f}$, and asteroseismic global parameters derived from the asymptotic
development. Temperature, luminosity and final metallicity are fixed in the code; all other parameters are free, including the mass and the age. The difference between the computed models and the observations is quantified by the $\chi^2$ function:
\begin{equation}
\chi^2 = \sum_{i=1}^N \left({p_i^{\hbox{\rm\small obs}}-p_i^{\hbox{\rm\small mod}} \over \sigma_i^
{\hbox{\rm\small obs}}}\right)^2
\label{chi2}
\end{equation}
Best-fit models are found by minimizing this function.  We then obtain an estimate of the parameters which cannot be measured by observations, such as the mass of the star, its age, the initial abundance of helium $Y\ind{i}$ and the initial metallicity $[$Z/X$]\ind{i}$.

The value of the initial helium content cannot be measured, but it can be assessed using the relative helium to metal enrichment of the galaxy $\Delta Y / \Delta Z$ through the relation:
\begin{equation}
Y={\Delta Y \over \Delta Z} \ Z+Y\ind{p}
\label{dcompos}
\end{equation}
where $Y\ind{p}$ is the primordial helium content. \cite{1998MNRAS.298..747P} have shown that
the enrichment was such that $2 \le \Delta Y / \Delta Z \le 5$. With a primordial helium abundance of $Y\ind{p}=0.238$, the initial helium content in the star is found between 0.25 and 0.27.

\subsection{Results}
\subsubsection{Taking into account the mean large spacing}

\begin{figure}
\centering
\includegraphics[width=8.5cm]{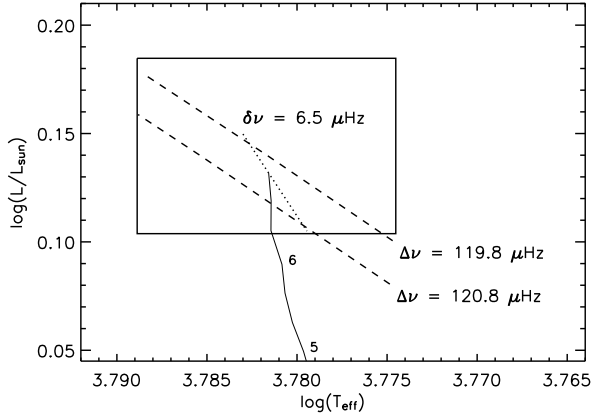}
\caption{Position in the HR diagram of the models which fit the mean value of the large spacing. The inner frame represents the 1-$\sigma$ limits of $\log(T\ind{eff})$ and $\log(L/L_{\odot})$. The dashed lines set the 1-$\sigma$ limits of the large splitting. The dotted line represents the location in the HR diagram of the models having a mean value of $\delta\nu_{02}=6.5 \mu$Hz, like the one derived from the observations. The solid line shows an example of an evolutionary track fitting $\Delta\nu$. Its characteristics are: $M=0.94 \ M_{\odot}$, age$ \ =6.9$ Gyr, $Y\ind{i}=0.26$, $[$Z/X$]\ind{i}=-0.38$. The numbers along the evolutionary track stand for the age of the star in Gyr.
\label{evoltrack}}
\end{figure}

The observed frequencies from the data analysis range from 2.3 mHz to 3.2 mHz. The large spacing derived from the observations is  determined with an uncertainty of 0.5\,$\mu$Hz in this frequency range, for the modes $\ell=0$, 1 and 2.
The first models had to fit the values of $T\ind{eff}$, $L$, $[$Z/X$]\ind{f}$, $[$Z/X$]\ind{i}$ and Y$\ind{i}$, plus this single asteroseismic constraint: only models which fit the mean large spacing with an accuracy better than 0.5\,$\mu$Hz were kept. The resulting position of the star in the HR diagram is narrowed (Fig.~\ref{evoltrack}) and the precision of the parameters of the star is improved (Table \ref{param-fit}).

\begin{figure}
\centering
\includegraphics[width=8.5cm]{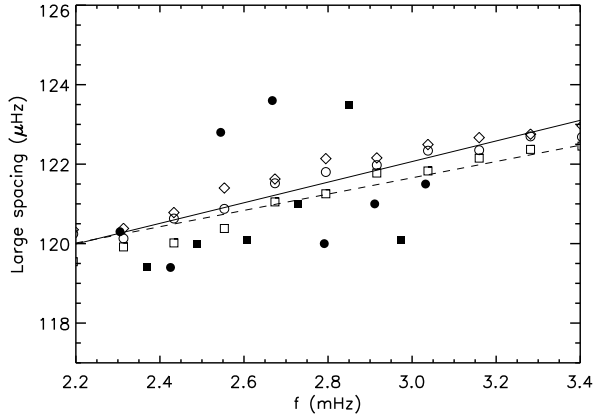}
\caption{Variation of the large spacing with frequency, for the $\ell$=0 (squares), 1 (circles) and 2 (diamonds) modes. The full symbols and the dashed line stand for the observations. The open symbols and the solid line stand for a model fitting the observational constraints.
\label{variationls}}
\end{figure}

\begin{figure}
\centering
\includegraphics[width=8.5cm]{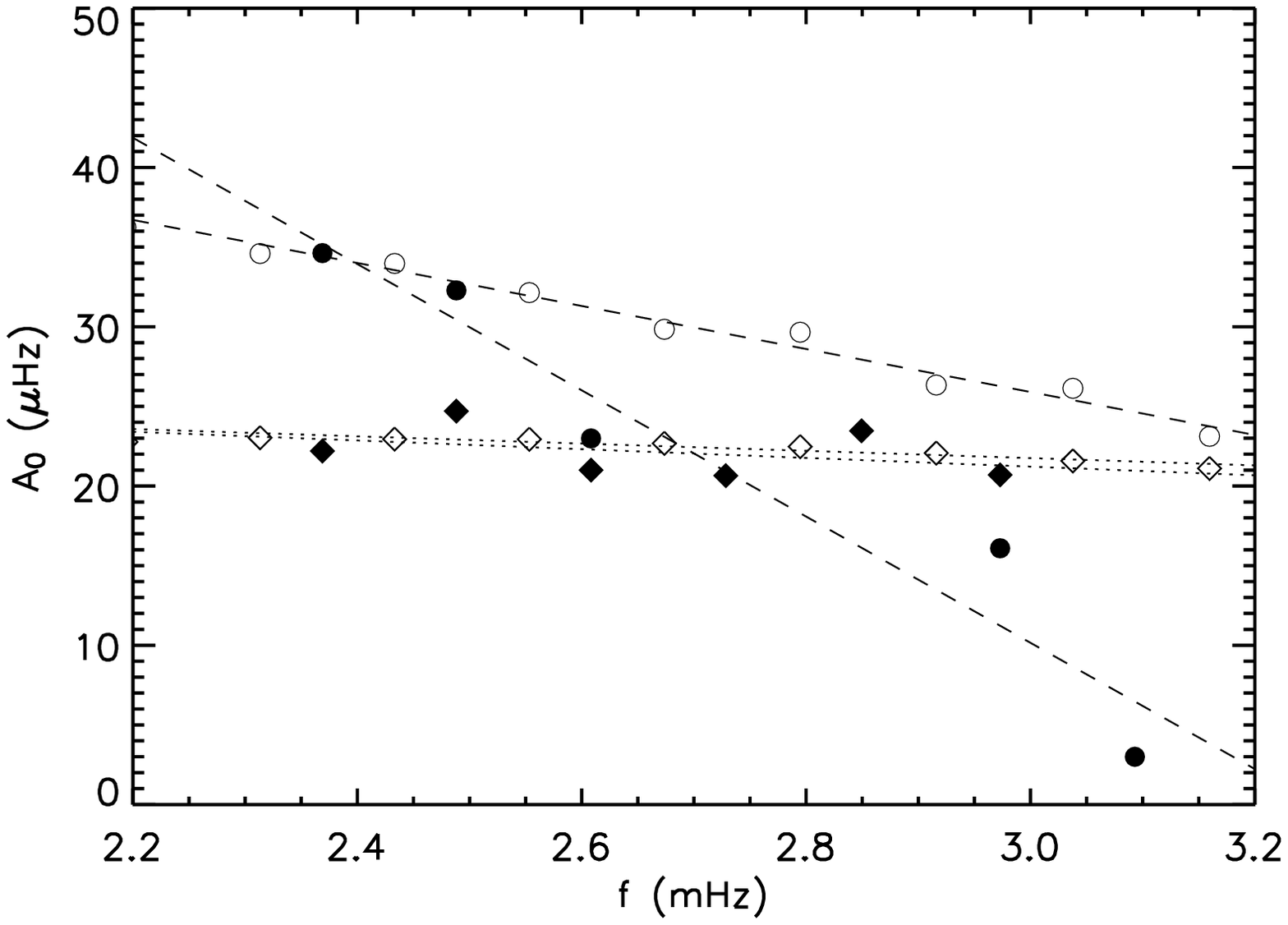}
\caption{Variation of the second order terms with frequency, derived from the $\ell=$ 1 (circles and dashed lines) and 2 (diamonds and dotted lines) modes, compared to $\ell=0$. The open symbols represent the model and the full symbols, the observations. Agreement is better in the case 2-0 than in the case 1-0, since the perturbation due to a dense core depends mainly on the factor $n+\ell/2$.
\label{variationA0}}
\end{figure}

\subsubsection{Taking into account the second order terms}

Taking into account the other asteroseismic parameters is required to constrain the models more efficiently. We therefore use the development expressed by \Eqt~\ref{provost}, which allows us to take into account the variation of the large spacing with frequency, and the different behaviors of the small spacings $\delta\nu_{01}$ and $\delta\nu_{02}$ (\Eqt s \ref{provost01} and \ref{provost02}). The observational asteroseismic constraints for the stellar modeling are then: $\Delta\nu$, $\courb$, $P_{1}$, $Q_{1}$, $P_{2}$ and $Q_{2}$.

Figure \ref{variationls} shows for instance the evolution of the large spacing with frequency, both for the observations and for one of the best-fit models. These evolutions are apparently in very good agreement: the values of the large spacing at 2.6 mHz agree within 0.1\% and the slope of the variations of the large spacing within 10\%. The only parameter which shows a marginal agreement is the one describing the evolution of $\delta\nu_{01}$, i.e. $Q_1$, as can be seen on Fig.~\ref{variationA0}. In fact, none of the models reaches the value of the slope derived from the observations. The agreement in the case 2-0 simply derives from the fact that the perturbation due to a dense core depends mainly on the factor $n+\ell/2$, and does not affect significantly the spacing $\delta\nu_{02}$. On the other hand, it strongly affects the spacing $\delta\nu_{01}$. Hence, a small discrepancy between the best fit model and the observation translates into a large discrepancy on this small spacing.

\begin{figure}
\centering
\includegraphics[width=8.5cm]{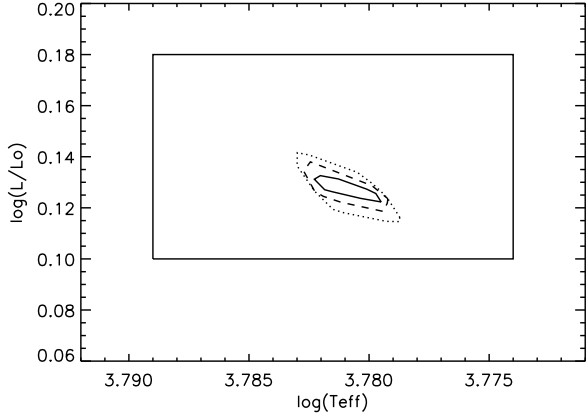}
\caption{Minimization of the $\chi^2$ function. The minimum value of $\chi^2$ is $\chi^2\ind{min} \simeq 9.5$. The solid line, the dashed line and the dotted line delimit the areas in the HR diagram at 1, 2 or 3 $\sigma$ where respectively $\chi^2 \le  \chi^2\ind{min}+1, 4$ and 9. As in Fig.~\ref{evoltrack}, the inner frame represents the 1-$\sigma$ uncertainties previous to this asteroseismic run.
\label{courbeniveaux}}
\end{figure}

This phenomenon mainly accounts for the impossibility to get very low values for $\chi^2$ (the minimum value is $\chi^2\ind{min} \simeq 9.5$).
These observational constraints yield an estimate of the stellar parameters, which are given in the second part of table \ref{param-fit}. Fig.~\ref{courbeniveaux} shows the position of \cible\  in the HR diagram resulting from the minimization.

Even if sensitive diffusion effects are expected for such a target, models were also computed without adding microscopic diffusivity, in order to study the impact on the evolution of the star. Agreement is worse, the best model providing $\chi^2\ind{min} \simeq 12.3$. The resulting models are expectedly older than those including diffusion (about 8 Gyr old), and the effective temperature is about 100\,K higher.

\begin{table}
\caption{Physical and seismic parameters of \cible\ derived from the modeling of the star (all models calculated with the convection parameter $\lambda = 0.94$). The
models of type \textit{a} are those which only take into account the mean value of the large spacing. The models of type \textit{b} are those which use the fitted values of $\Delta\nu$, $\courb$, $[$Z/X$]\ind{i}$, $P_{1}$, $Q_{1}$, $P_{2}$ and $Q_{2}$.
\label{param-fit}}
\begin{tabular}{lrr}
\hline
Physical parameters & models \textit{a} & models \textit{b} \\
\hline
$T\ind{eff}$    & 6050 $\pm$ 100 K  & 6037 $\pm$ 19 K\\
$L/L_\odot$     & 1.39 $\pm$ 0.13   & 1.341 $\pm$ 0.016\\
$[$Z/X$]\ind{f}$& $-0.55\pm 0.05$ dex& $-0.54 \pm 0.03$ dex\\
\hline
$Y\ind{i}$      & 0.26 $\pm$ 0.01   & 0.252 $\pm$ 0.003\\
$[$Z/X$]\ind{i}$& $-0.42\pm 0.05$ dex& $-0.416 \pm 0.03$ dex\\
$M/M_\odot$     & 0.93 $\pm$ 0.07   & 0.939 $\pm$ 0.009 \\
age             & 7.5  $\pm$ 2.6 Gyr& 7.25 $\pm$ 0.07 Gyr\\
$R/R_\odot$     & & 1.064 $\pm$ 0.014\\
\hline
Seismic parameters& models \textit{a}$^\flat$ & models
\textit{b}$^\sharp$ \\
\hline
$\Delta\nu$ & $120.3 \pm 0.5\,\mu$Hz & $120.4 \pm 0.4\,\mu$Hz\\
$\delta\nu_{02}$ & $6.9 \pm 1.5\,\mu$Hz & $6.75 \pm 0.05\,\mu$Hz\\
$\courb$   &   & 0.290 $\pm$ 0.003\,$\mu$Hz\\
$P_1$   &  & $32.54 \pm 0.48\, \mu$Hz\\
$P_2$   &  & $22.50 \pm 0.14\, \mu$Hz\\
$Q_1$   &  & $-1.78 \pm 0.06\, \mu$Hz\\
$Q_2$   &  & $-0.33 \pm  0.05\, \mu$Hz\\
\hline
\end{tabular}
$^\flat$ mean value over the frequency range [2.3 - 3.2 mHz]\\
$^\sharp$ at 2.6 mHz
\end{table}

Finally, our study yields to a star reaching the end of the main sequence. The hydrogen is almost entirely exhausted in the center: the mass fraction of remaining hydrogen in the core is of about 13\%. The best models show no trace of a convective core during the evolution sequence, except plausibly during the first billion of years.
As explicited by Tables \ref{prop-phys} and \ref{param-fit}, the precision on the stellar parameters has been significantly improved. Localization in the HR diagram is refined by a factor greater than 5 in temperature and about 8 in luminosity. The error bar on the mass is now defined, as low as 0.01\,$M_\odot$. The precision on the age, is much better, and we note that the age of the star corresponds to the low limit given by previous works (6.5$\to$14.5\,Gyr according to \cite{2006ApJ...636.1098B}).

\section{Conclusion\label{conclusion}}

This single-site 5-day long asteroseismic run on \cible\ has given a much more precise view of this star of the old galactic disk.
We have developed a method for extracting the peaks with the lowest false alarm probability. This method proves to be efficient for single-site observation with rapidly varying photometric conditions. The performing of multiple periodograms combined with a statistical test allows us to extract more peaks than with a single treatment. Using a criterion combining the minimization of false alarm probabilities and \'echelle diagram analysis, we have identified \nident\ $\ell=0$ and $\ell=1$ eigenmodes in the spectrum of \cible, from which we have derived lower amplitudes modes (including $\ell=2$ modes) as well as the asymptotic parameters.

Despite the very short duration of the run, yielding a limited precision for the identified eigenfrequencies, the fitting of the spectrum has required a more precise development than the usual second order term $-\ell(\ell+1)\, D_0$. We have shown that the Tassoul original form with a second order term decreasing with frequency $-\ell(\ell+1)/(n+\ell/2)\, A_0$ must be preferred to the development $-\ell(\ell+1)\, D_0$ often used for interpreting ground-based observations. Nevertheless, the asymptotic development cannot account for the precise oscillation pattern of \cible : this star exhibits clearly a large spacing with a sensitive dependence on the radial order, and small spacings depending on the mode degree. This dependence observed in the data are confirmed in the modeling: the strong composition and sound speed gradients in the small core mainly composed of helium are responsible of the modulation of the oscillation pattern. The eigenfrequency precision in our data set is not accurate enough to give additional independent constraints on the core size; continuous long-duration observations are required for such a task.

Contrary to a similar F type dwarf target (HD 49933, \cite{2005A&A...431L..13M}), \cible\ does not exhibit any noticeable activity. This may be due to the geometric configuration of the observation, with a possible very low inclination axis. By now, \cible\ presents the lowest metallicity among dwarfs observed in asteroseismology. Similarly to HD 49933, modes amplitudes are sensitively smaller than expected from the scaling law. Two effects may explain this: first, both stars are undermetallic; second, the scaling has not yet been calibrated on dwarf F stars. Observations with the satellite CoRoT will help understanding that behavior.

The modeling of \cible\ has been achieved with the evolution code CESAM2k (\cite{1997A&AS..124..597M}). Taking into account the asteroseismic constraints (large and small spacings) allows us to propose a much more precise description of this star. Error bars on the physical parameters have been divided by a factor of 2 to 8, in the framework of the present physical description used in this work. The age we derive for \cible\ is about 7.25$\pm$0.07\, Gyr. Hydrogen is almost exhausted in the core. With improved values of $L$, $T\ind{eff}$ and $\log g$ (4.356$\pm$0.016), a better signature of the metallic abundances is certainly possible and may help to improve the localization of this star in the thin or thick galactic disk.



\end{document}